\documentclass{elsart}

\usepackage{amssymb}

\begin{document}

\begin{frontmatter}



\title{On amplitude oscillation of vibrations of strongly anisotropic
high-temperature superconductors of BiPbSrCaCuO system.}


\author{J. G.  Chigvinadze,  A. A. Iashvili, T. V. Machaidze}

\address{E. Andronikashvili Institute of Physics, 380077 Tbilisi, Georgia}

\begin{abstract}
Effect of oscillations of the vibration amplitude of cylindrical
sample suspended by a thin elastic thread and vibrating in a
transverse magnetic field and containing $2D$
quasi-two-dimensional vortices (pancakes), was observed in the
strongly anisotropic high-$T_{c}$ superconductor of
$Bi_{1.7}Pb_{0.3}Sr_{2}Ca_{2}Cu_{3}O_{y}$ system.
\end{abstract}
\begin{keyword}
Pinning \sep Vortex lattice

\PACS 74.60.-w \sep 74.60.Ge \sep 74.80.Dm
\end{keyword}
\end{frontmatter}

At investigation of $3D-2D$ transition in anisotropic
high-temperature superconductors [1], we used the supersensitive
mechanical method of investigation of dissipation processes [2].
Measuring the logarithmic decrement of damping $\delta$ as a
function of outer magnetic field in the mixed state, a sharp
decrease of dissipation was established which finished above the
determined values of the field (of the order of 2000 Oe). This
decrease is so great that the dissipation turns to be equal to
that of background observed without magnetic field. This large
change of dissipation, apparently, is connected with the increase
of critical current, which should be observed at the transition of
$3D$ three-dimensional vortices in quasi-two-dimensional $2D$ ones
[3].

At a further increase of magnetic field the dissipation is not
practically changed, in any case, at fields of the order of 2000
Oe and higher up to some definite field on the curve of magnetic
field $H$ dependence of $\delta$ the plateau is observed in some
case, but at higher values of $H$ the dissipation begins to
decrease slightly and sometimes it turns out to be lower then its
value without magnetic field (naturally, the sample temperature is
$T<T_{c}$).

Let us present, for example, the $\delta=f(H)$ dependence at
temperature $T=93$ K (Fig.1). The critical temperature of our
sample is $T_{c}=107$ K.

As it is seen from the figure, the plateau of logarithmic
decrement of damping is observed on $\delta=f(H)$ curve in the
range of magnetic field from 2000 to 2800 Oe. At further increase
of the magnetic field up to value $H\approx 5000$ Oe the energy
dissipation continues to decrease gradually.

The particular attention should be paid to the range of
dissipation plateau ($2000-2800$ Oe). In this range of magnetic
fields the sharp change of energy dissipation dynamics takes
place.  On the other hand, in magnetic field range from $H=0$ to
$H=2000$ Oe and from $H=2800$ Oe and higher, up to maximum field
$H\approx 5000$ Oe measured by us, the character of dissipation is
standard: the vibration amplitude decreases gradually by
exponential law (see Fig.2) and from these data $\delta$ is
defined. In contrast to this picture in the range of fields from
$H=2000$ Oe to $H=2800$ Oe the time dependence of vibration
amplitude is sharply changed and pronounced oscillations with the
period of $20\div70$ sec. occur disappearing just as soon as the
magnetic field value leaves the plateau region. In the given case
at temperature $T=93$ K the oscillations disappear and
$\delta=f(H)$ dependence turns out to be as smooth as at fields
$H<2000$ Oe.

On $\delta=f(H)$ curve plateau is not always observed at fields
more then $H>2000$ Oe at higher temperatures  but there is a
gradual decrease of vibration damping. As an example of such
behavior let us take $\delta=f(H)$ dependence for temperature
$T=100$ K (Fig.3). Here, also as at $T=93$ K, when the magnetic
field becomes $>2000$ Oe, in particular, when fields $H>3000$ Oe,
the vibration amplitude oscillations appear (Fig.4), disappearing
at $H=5300$ Oe. It should be noted that the frequency of
suspension system also oscillates with the amplitude and one
should especially emphasize that maximum of vibration amplitude
coincides with the frequency maximum, and the minimum of vibration
amplitude with the minimum of frequency. One of possible
explanations of this phenomenon, in particular, the oscillation of
frequency and amplitude, is that in the crystal lattice of
investigated sample there are directions along which vortices
could be fixed more effectively than along other directions.

As it has been already noted, in $2000-2800$ Oe interval the time
dependence of amplitude is not exponential and as a result, the
usual measurement of dissipation decrement becomes impossible. The
values for this range shown in Fig. 1 and Fig. 3 are determined by
envelops of curves, shown in Fig.4a and Fig.4b.

After the appearance of the oscillations of vibration amplitude
their frequency first decreases with the increase of magnetic
field, and the amplitude increases. The amplitude increase and
frequency decrease process continues up to the definite field and
further, on the contrary, the amplitude decreases with the
increase of field, and the frequency of oscillations decreases
and, finally, at $H=5300$ Oe oscillations disappear. At fields
$H>5300$ Oe the dependence of amplitude on time turns to be as
smooth as at fields $H<2000$ Oe, shown in fig.2.

It turns out that the similar behavior of the frequency of
oscillations can be observed at a fixed field. With the decrease
of the amplitude of vibrations at $T=100$ K and $H=4200$ Oe, as it
is seen from the plot presented on Fig.4(a), the frequency of
amplitude oscillations first decreases, reaches the minimum and
then it begins to increase, with the decrease of amplitude, down
to the full disappearance of oscillations.

The investigation of temperature and field dependence of the
beginning and ending of oscillations gave us the possibility to
reveal the region on $H(T)$ diagram where the oscillations of
vibration amplitude are observed - Fig.5 (dashed region). As it is
seen from the figure, with the decrease of temperature this region
of field is narrowed and it is possible that at low temperatures
it can even disappear.

In the previous work [1] the onset of plateau we related to the
final decay of three-dimensional $3D$ Abrikosov vortices and
appearance of quasi-two-dimensional $2D$ vortices (pancakes). At
the same time we proceeded from the considerations given in work
[3], where with the help of computer calculation it was shown that
at the decay of $3D$ vortices into quasi-two-dimensional $2D$
ones, the critical current should increase drastically. This means
that the dissipation is decreased as it was observed by us.

Note that the estimation of the field value at which $2D$
quasi-two-dimensional vortices (pancakes) begin to appear gives
namely field $H\approx 2000$ Oe [1,4].

One more argument in favor for the $3D-2D$ transition at field
$H\approx2000$ Oe is related to the fact that the amplitude
effects are strongly suppressed at fields higher than the
transition to quasi-two-dimensional state (at fields $\geq2000$
Oe). Both the suspension system vibration frequency, and the
damping in these fields and in higher ones depend more weakly on
amplitude (Fig.6), than at weak fields when the vortices are
three-dimensional $3D$ ones. However, as for   the amplitude
dependence of damping and vibration frequency for the regions
where amplitude oscillations are observed, if they are plotted to
the same scale for the magnetic field region with $2D$ vortices
$(H>2000$ Oe) (Fig.6) and weak magnetic fields where $3D$
three-dimensional vortices are observed $(H<<2000 $ Oe) (Fig.7),
it is seen that in the case of quasi-two-dimensional $2D$ vortices
in the amplitude oscillation region amplitude effects are weaker
than in the case of $3D$ vortices. However, under these conditions
the amplitude effects are smaller as compared with the great
amplitude effects observed in classical superconductors of the
second kind for three-dimensional $3D$ Abrikosov vortices [2,5].

In the case of field $H=110$ Oe (Fig.7), vortices are clearly
three-dimensional Abrikosov ones, though in these layered samples
of high-temperature BiPbSrCaCuO superconductor these amplitude
effects are not so strong as in classical superconductors of the
second type, but, they clearly denote the significant amplitude
dependence of both $\delta$ and $\omega$. As it is seen from the
presented plot, here the amplitude effects are higher than their
values in fields being close to and higher than $3D-2D$ transition
fields. For classical superconductors of II type, where the
amplitude effects are considerably larger, the data were presented
in works [2,5]. The amplitude effects caused by the motion of
Abrikosov vortices in the vibration experiments with
superconductors of II type where observed by us for the first time
[6,7] and explained by V. Galaiko [8] who connected these effects
with concentrations of pinned and free vortices. According to his
theory the concentrations of free and pinned vortices change with
the change of magnetic field and their interaction is the reason
of damping and, consequently, the vibration damping in the mixed
state changes with the change of magnetic field. The similar
amplitude effects for $\delta$ and $\omega$ were observed in
high-temperature superconductors as well [9,10].

The significant decrease of amplitude effects in the layered
high-temperature superconductors at fields $H>2000$ Oe points to
the fact that in our experiments the vibration amplitudes are not
sufficient for tearing of vortices from pinning-centers and
therefore, apparently, practically all vortices are pinned, and
the concentration of free vortices is almost equal to zero
$n_{fr}\approx 0$ and the concentration of fixed vortices is
almost equal to unit $n_{pin} \approx 1$, resulting to the
significant decrease of dissipation energy. However, it should be
noted that, apparently, a small concentration of free vortices
remains different from zero $n_{fr}\neq 0$ causing the oscillation
effects of vibration amplitude, and also the dissipation processes
at fields $H>2000$ Oe being too small, but still different from
zero.

The very fact of the existence of vibration amplitude oscillation
is the most unclear problem. It is not clear where the energy for
vibration amplitude enhancement comes from and why its
appearing-disappearing has a periodical character.

Possible doubt that this effect, probably, is not connected with
the processes taking place in the investigated superconductors,
but it is caused by superposition of different vibrations on
axial-and-torsional vibrations of the cylindrical sample (the
description of the device installation is given in [1]) is
verified by the following facts:

1). The frequency of intrinsic axial vibration of the system
$(\omega_{0}=1,2560$ sec$^{-1}$) is not in simple relation with
the frequency of radial vibrations $\omega_{r}=9,96825$
sec$^{-1}$; 2) Using the other sample with considerably smaller
number of vortices, the mechanical vibration frequencies at zero
field remained unchanged, but frequency of axial vibrations at
magnetic field in the range of vibration amplitude oscillation was
noticeably changed (from $\omega\approx 3,2\div4,3$ sec$^{-1}$ to
$\omega\approx 2$ sec$^{-1}$).

The testing experiment eliminated also the possibility of the
influence of mechanical noise in vibrating system, its inertia
moment (by replacement of discus) and of the change of torque (by
replacement of elastic thread by two - upper and lover threads
[11], intrinsic frequency ($\omega_{0}=1,047$ sec$^{-1}$) and
frequencies at the presence of field $\omega=1,5\div1,75$
sec$^{-1}$ were essentially changed, but the magnetic field
interval, where the oscillation of vibrations is observed,
remained unchanged).

In the testing experiment of the other type the inertia moment was
continuously changed almost four times in the process of
experiment. These experiments will be described in more detail in
the next paper.

These experiments showed that in spite of the change of intrinsic
frequency of vibration system, the vibration amplitude
oscillations are observed at the same magnetic field interval, as
in the above described experiment.

Thus, we made sure that the phenomena of vibration amplitude
oscillations are directly connected with processes taking place in
superconductors being in magnetic field, rather than with side
effects.

Let us try to explain situation at lest in general terms.

As it is know [12,13] in the tilted field when there are weak
Josephson links between the layers, the quasi-two-dimensional $2D$
system of vortices is created, interconnected by Josephson
junctions (Josephson links) that is realized only in the case when
the angle of inclination of internal field 
$\theta_{h}>\theta_{0}=arctg\Gamma$, where $\Gamma$ is the
anisotropy factor $\Gamma=M/m$, $M$ is the effective mass along c
axis, and $m$ is the effective mass in the basis $(a,b)$ plane. In
our case $\Gamma=3000$ and the estimation gives $\theta_{0}\approx
89,98^{\circ}$ .

The angles of rotation in our experiment not exceed
$\varphi\approx1,0$ rad. Consequently, in our experiments the
Josephson vortices do not appear the creation and disintegration
of which could, principally, cause the phenomenon of vibration
amplitude oscillation. Oscillations can not be also explained
either by of R. A. Klemm's model [14]. The mechanical moment
oscillations predicted by him should be observed at large angles
$\theta>\theta_{cr} =89,64^{\circ}$, being more than the maximal
values of vibration amplitude in our experiments
$\varphi_{max}\approx1,0$ rad.

One of the possible mechanisms of phenomenon observed by us can be
connected with the following: during vibration of our sample $2D$
vortices together with the sample are turned relative to the
magnetic field and, as a result, the flux of magnetic field in
pancakes planes parallel to the basic plane of the crystal
(\textbf{a},\textbf{b}) is changed. Thus, the alternative
electromotive force of induction arise causing the creation of
alternative field superimposed on the basic steady magnetic field
$H$. The interaction of magnetic field with the sample takes
place.

The phase shift between forcing vibrations (axial vibrations)
being possible in this case and forced vibrations (pancake
vibrations) can be the cause of amplitude oscillation, as it
increases at the coincidence and closeness of phases but with the
divergence of phases the amplitude increase is changed by its
decrease.

So far we have not been able to give the more detailed description
of this process which, apparently, should be connected with the
vortex slipping (vortices are more strongly pinned in the
direction perpendicular to the direction of their vibrations than
in the direction of vibrations).

One should also discuss the possibility of generation of vortices
of the different waves and their possible influence on the
character of the sample vibrations in the system investigated by
us.

This work is made with support of International Scientific and
Technology Center (ISTC) through Grant  G-389.




\newpage
\begin{center}
{\large Figure Captions}
\end{center}
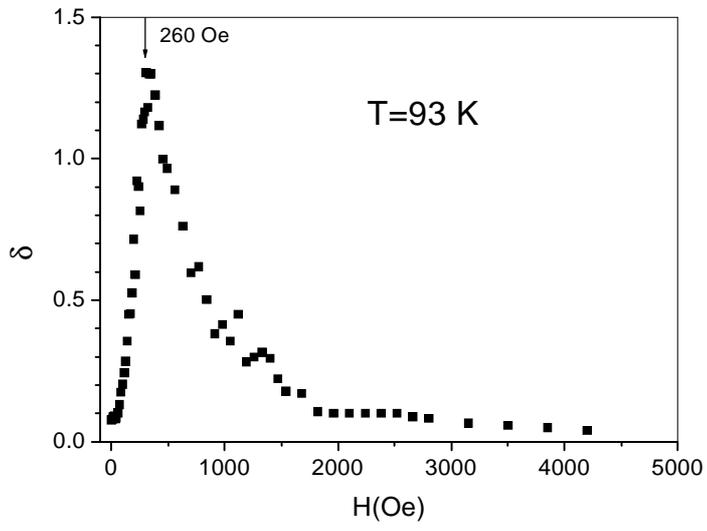
Fig.1. Logarithmic decrement of damping $\delta$ dependence on the
strength of outer magnetic field $H$. Temperature $T=93$ K.

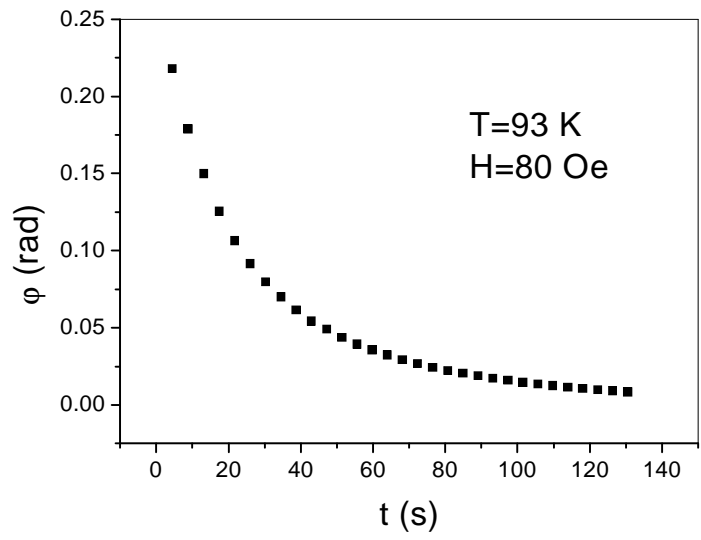
Fig.2. Vibration amplitude decrease of the suspension system on
time. $T=93$ K, $H=80$ Oe.

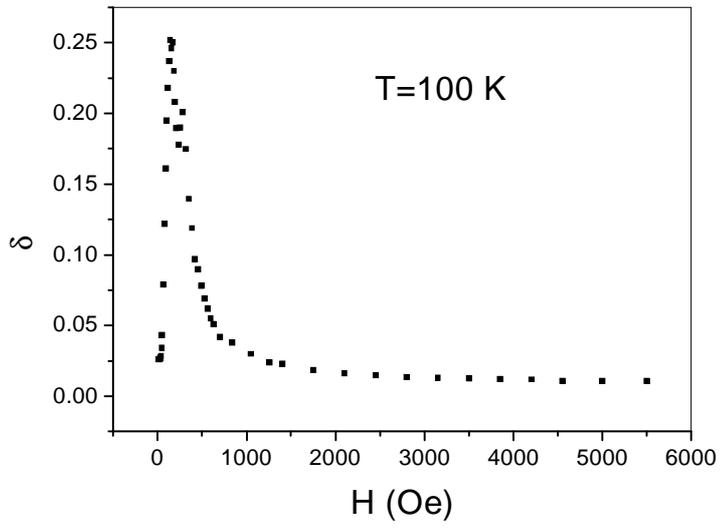
Fig.3. Logarithmic decrement of damping $\delta$ dependence on the
strength of outer magnetic field $H$. Temperature $T=100$ K.

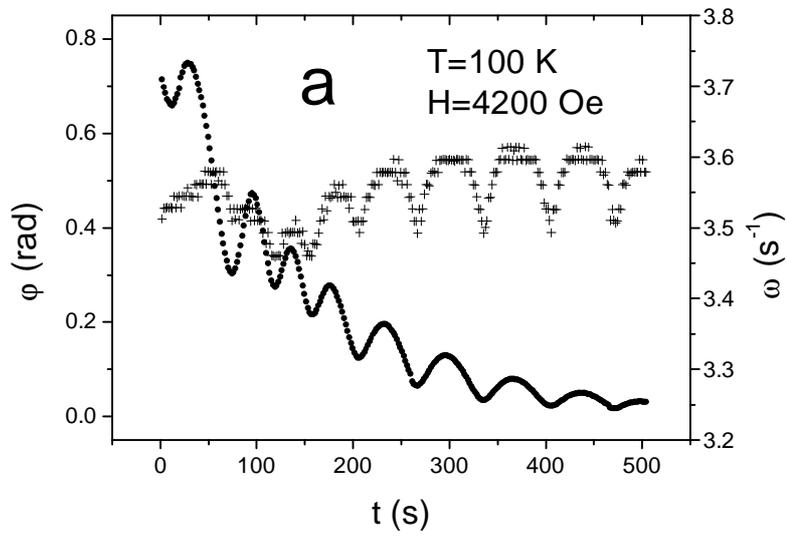
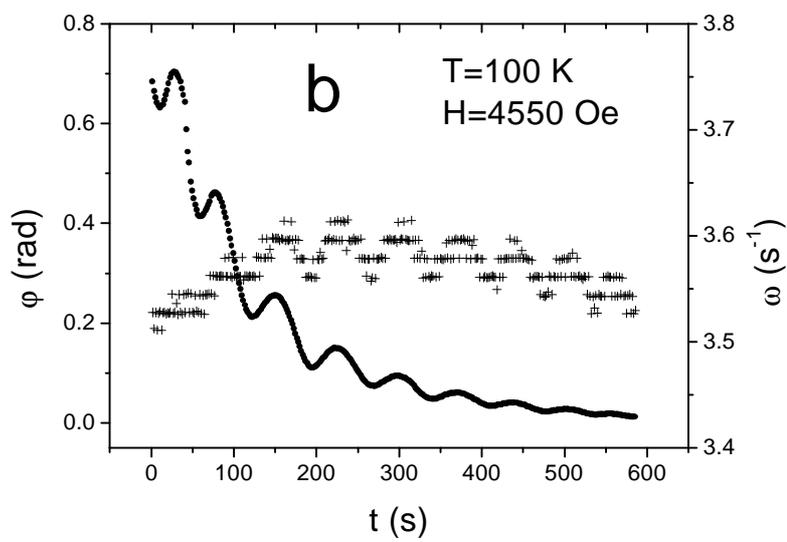
Fig.4. Vibration amplitude and frequency dependence of the
suspension system on time. $T=100$ K, $\textbf{a}-H=4200$ Oe,
$\textbf{b}-H=4550$ Oe.

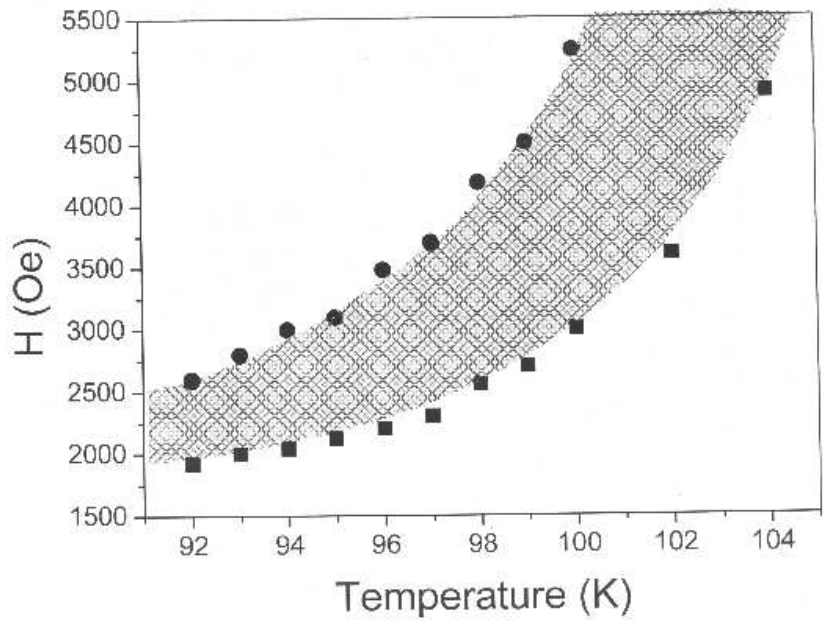
Fig.5. $H(T)$ phase diagram. Here in the dashed part of diagram
oscillations of vibration amplitude of the suspension system are
observed.

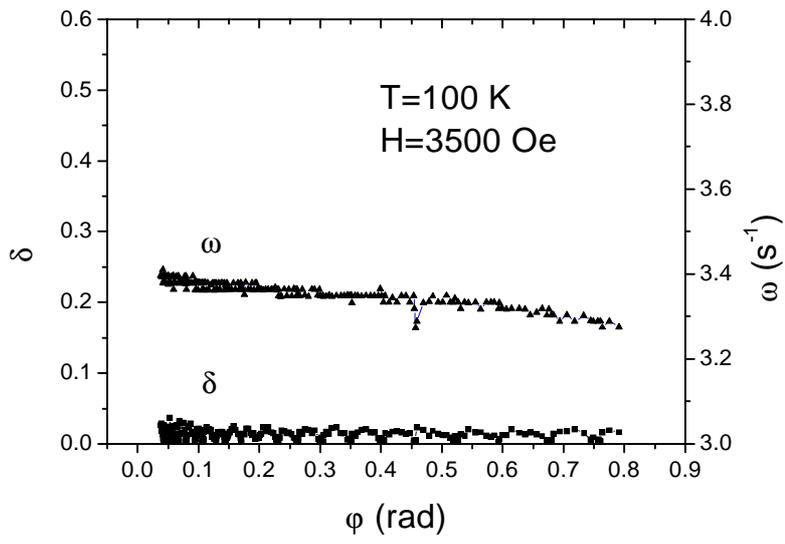
Fig.6. Damping $\delta$ and vibration frequency $\omega$
dependence on amplitude. $T=100$ K, $H=3500$ Oe.

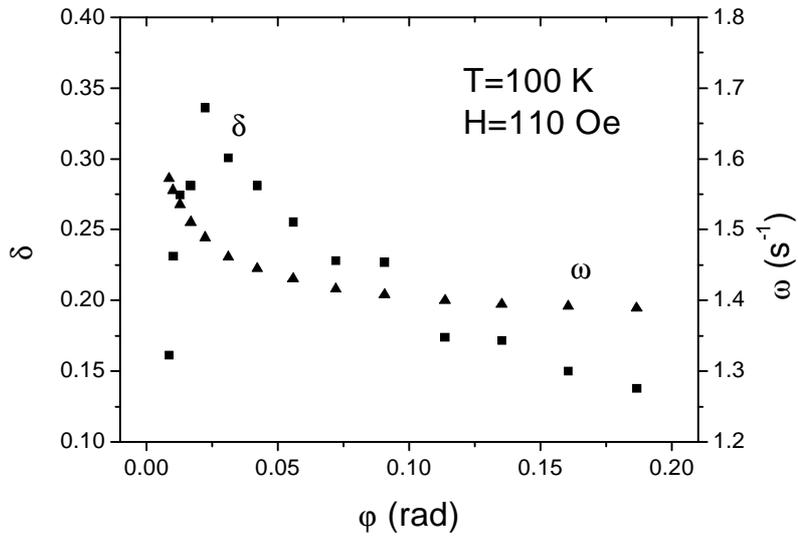
Fig.7. Damping $\delta$ and vibration frequency $\omega$
dependence on amplitude. $T=100$ K, $H=110$ Oe.

\end{document}